\documentclass{aa} 

\usepackage{amsfonts}
\usepackage{amsmath}
\usepackage{amssymb}
\usepackage{graphicx}
\usepackage{gensymb}
\usepackage{listings}

\usepackage{lscape}
\usepackage{subfigure}
\usepackage{natbib}
\usepackage{hyperref}
\usepackage[varg]{txfonts}
\usepackage{longtable}
\usepackage{booktabs}
\usepackage[flushleft]{threeparttable}

\usepackage{xcolor}

\bibpunct{(}{)}{;}{a}{}{,} 

\begin{document}

   \title{Extending the variability selection of active galactic nuclei in the W-CDF-S and SERVS/SWIRE region \thanks{Table \ref{tab::catalog} is only available in electronic form at the CDS via anonymous ftp to cdsarc.u-strasbg.fr (130.79.128.5) or via \url{http://cdsweb.u-strasbg.fr/cgi-bin/qcat?J/A+A/}.}}

   \author{M. Poulain\inst{1}, M. Paolillo\inst{2,3,4}, D. De Cicco\inst{5,6,2}, W. N. Brandt\inst{7,8}, F. E. Bauer\inst{6,5,9},  S. Falocco\inst{10}, F. Vagnetti\inst{11}, A. Grado\inst{4}, F. Ragosta\inst{2,4,3}, M. T. Botticella\inst{4}, E. Cappellaro\inst{4}, G. Pignata\inst{12,5}, M. Vaccari\inst{13,14}, P. Schipani\inst{4}, G. Covone\inst{2,3}, G. Longo\inst{2,3,4}, N. R. Napolitano\inst{15}}
   \titlerunning{Optically variable AGN in the VST/CDF-S}
   \authorrunning{M. Poulain et al.}
\institute{
Institut f\"{u}r Astro- und Teilchenphysik, Universit\"{a}t Innsbruck, Technikerstra\ss e 25/8, Innsbruck, A-6020, Austria 
\\e-mail: melina.poulain45@gmail.com, Melina.Poulain@uibk.ac.at
\and
Department of Physics, University of Napoli ``Federico II'', via Cinthia 9, 80126 Napoli, Italy
\and
INFN - Sezione di Napoli, via Cinthia 9, 80126 Napoli, Italy 
\and
INAF - Osservatorio Astronomico di Capodimonte, via Moiariello 16, 80131 Napoli, Italy 
\and
Millennium Institute of Astrophysics (MAS), Nuncio Monse\~nor Sotero Sanz 100, Providencia, Santiago, Chile    
\and 
Instituto de Astrof\'{i}sica, Pontificia Universidad Cat\'{o}lica de Chile, Av. Vicu\~{n}a Mackenna 4860, 7820436 Macul, Santiago, Chile 
\and
Department of Astronomy and Astrophysics, The Pennsylvania State University, University Park, PA 16802, USA 
\and
Institute for Gravitation and the Cosmos, The Pennsylvania State University, University Park, PA 16802, USA        
\and
Space Science Institute, 4750 Walnut Street, Suite 2015, Boulder, CO 80301, USA 
\and
KTH Royal Institute of Technology, Brinellv\"{a}gen 8, 114 28 Stockholm, Sweden 
\and
Department of Physics, University of Roma ``Tor Vergata'', via della Ricerca Scientifica 1, 00133 Roma, Italy 
\and
Departamento de Ciencias Fisicas, Universidad Andres Bello, Avda. Republica 252, Santiago, Chile 
\and
Department of Physics and Astronomy, University of the Western Cape, Private Bag X17, 7535 Bellville, Cape Town, South Africa      
\and
INAF - Istituto di Radioastronomia, via Gobetti 101, 40129 Bologna, Italy   
\and
School of Physics and Astronomy, Sun Yat-sen University, Guangzhou 519082, Zhuhai Campus, P.R. China    
}

\date{}

\abstract
{Variability has proven to be a powerful tool to detect active galactic nuclei (AGN) in multi-epoch surveys. The new-generation facilities expected to become operational in the next few years will mark a new era in time-domain astronomy and their wide-field multi-epoch campaigns will favor extensive variability studies.}
{We present our analysis of AGN variability in the second half of the VST survey of the Wide Chandra Deep Field South (W-CDF-S), performed in the \emph{r} band and covering a 2 sq. deg. area. The analysis complements a previous work, in which the first half of the area was investigated. We provide a reliable catalog of variable AGN candidates, which will be critical targets in future variability studies.}
{We selected a sample of optically variable sources and made use of infrared data from the \emph{Spitzer} mission to validate their nature by means of color-based diagnostics.}
{We obtain a sample of 782 AGN candidates among which 12 are classified as supernovae, 54 as stars, and 232 as AGN. We estimate a contamination $\lesssim 20\%$ and a completeness $\sim 38\%$ with respect to mid-infrared selected samples. }
{}

\maketitle
   
\section{Introduction}
\label{section:introduction}
The role of active galactic nuclei (AGN) in astrophysics and cosmology has increased in recent decades owing to the discovery that their properties correlate with those of the host galaxies \citep{Magorrian, Kormendy} and to the star formation rate \citep{Mullaney, Aird}. This implies that the evolution of large-scale structures in the Universe is closely linked and possibly regulated, in part, by the nuclear activity in the core of galaxies. Most cosmological simulations now include some form of active feedback from AGN in order to regulate star formation \citep[e.g.,][and references therein]{sijacki, booth} and evidence of AGN feedback in galaxies and clusters is rapidly growing \citep[e.g.,][]{fabian, gitti}. 
This implies that a complete census of AGN is mandatory to properly understand the evolution of cosmic structures.

Historically, many  techniques have been devised to identify AGN activity. Color diagnostics and spectra energy distribution (SED) fitting are some of the most common due to the increasing availability of multiwavelength observations over large areas of the sky from the ultraviolet (UV) to the infrared (IR). On the other hand, X-rays observations remain one of the most efficient ways to detect nonthermal radiation from galaxy nuclei, but unfortunately these observations still lack sensitive sky coverage comparable with those available at other wavelengths \citep{BrandtAlexander}.

One of the most general properties of AGN is their variability at every wavelength and timescales (from hours to years; see, e.g., \citealt{Ulrich}); therefore, an alternative way to find AGN is through the search for variable sources. This method however requires a large investment in observing time organized in extended monitoring campaigns, and these could often only be afforded with the use of smaller telescopes or focusing the analysis on small regions of the sky \citep[see, e.g.,][]{Trevese,Villforth,Sarajedini,Choi,Graham,Cartier,Simm,Pouliasis}.

In the last few years our group has dedicated its efforts to assessing the performance of AGN selection through variability down to the limits that are expected to be reached by next generation monitoring surveys such as those that will be conducted with the Large Synoptic Survey Telescope \citep[LSST; see, e.g.,][]{lsst}. To this end, we are exploiting the data collected by the Very Large Telescope Survey Telescope (VST) on some of the fields that are expected to be targeted by LSST in its Deep Drilling Field campaign \citep[e.g.,][]{LSST_WP}. In particular in \citet{DeCicco} and \citet{decicco19} we focused on the COSMOS field because of its extensive X-ray, IR, optical, and spectrocopic coverage, assembling a catalog of variability selected AGN; the three-year campaign allowed us to assess how variability selection compares to other selection techniques along with its dependence on baseline and sampling cadence. In addition in \citet{Falocco} we started analyzing VST data covering two square degrees in the Spitzer Extragalactic Representative Volume Survey \citep[SERVS;][]{Mauduit} and the Spitzer Wide-Area Infrared Extragalactic Survey \citep[SWIRE;][]{Lonsdale} region centered on the Extended Chandra Deep Field South (ECDFS). This area, labeled Wide-Chandra Deep Field-South (W-CDF-S), is of great interest since it is one of the approved LSST Deep Drilling Fields as well as one of the Euclid Deep Survey Fields, and the scientific community is making an effort to acquire as much ancillary data as possible before the beginning of LSST and Euclid activities \citep{W-CDF-S_opt,XMM-LSS,LSST_WP}. The present work is intended to extend AGN variability selection to an additional two square degrees to obtain an AGN sample covering the full central four square degrees of the W-CDF-S area to exploit the existing Spitzer and future X-ray coverage maximally (Figure \ref{fig::CDF-S}).

This paper is organized as follows. In Section \ref{Data} we present the VST telescope, the SUDARE-VOICE survey, and the dataset. In Section \ref{Catalog_extraction}, we describe the extraction process of the source catalog. In Section \ref{Selection} we explain how we selected the variable sources, i.e., our sample of AGN candidates. In Section \ref{Nature}, we describe the diagnostics we used to determine the nature of our AGN candidates. Finally, we discuss our results and draw our conclusions in Section \ref{Discussion_conclusion}.  

\section{Dataset}
\label{Data}
The data used for this work come from the 2.6 m VST telescope \citep{Capaccioli11}, located at Cerro Paranal Observatory in Chile. The VST is equipped with the OmegaCAM camera \citep{Kuijken}, consisting of 32 CCDs with a field of view (FoV) of 1\degree$\times$1\degree\ and a spatial resolution of $0.21\arcsec$/pixel. 

In this work we make use of data from the SUDARE-VOICE survey. The SUDARE \citep{Cappellaro} program is dedicated to the study of supernovae (SNe) in the COSMOS and Chandra Deep Field South (CDF-S) regions. The VOICE survey \citep{Vaccari16} meanwhile targets the CDF-S and ELAIS S1 regions. Both of these surveys obtained imaging in the optical \emph{g} and \emph{r} bands; additional observations were carried out in the \emph{i} band for SUDARE and in the UV \emph{u} band for VOICE. In this work we focus only on the \emph{r}-band data because they have the highest observing cadence, i.e., three days. For the other bands, observations were obtained every ten days.

\begin{figure}[t]
\begin{center}
        \includegraphics[width=\linewidth]{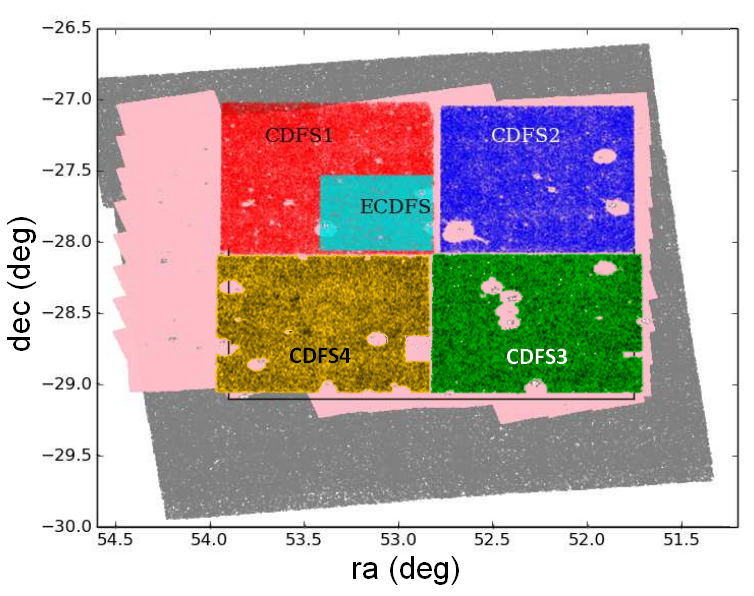}
        \caption{\label{fig::CDF-S}Chandra Deep Field South as imaged by different surveys (gray: SWIRE, pink: SERVS, cyan: ECDFS, other colors: VST-CDF-S regions).        Adapted from \citet{Falocco}. }
\end{center}
\end{figure}

The VST observations of the CDF-S cover four regions of one square degree each, which were imaged at different times: half the total area (CDF-S1 and 2) was analyzed in \citet{Falocco}, while this work focuses on the other half (CDF-S3 and 4; see Figure \ref{fig::CDF-S}). The dataset for each of the four areas is composed of several epochs. An epoch is obtained as the combination of five exposures of six minutes each, meaning that we have 30 minutes total exposure per epoch. In the case of CDF-S3, the dataset consists of 32 epochs distributed over one year and 11 months; seeing is characterized by full width at half maximum (FWHM) in the range $0.52 - 1.2\arcsec$. The CDF-S4 dataset consists of 30 epochs distributed over one year and four months; seeing FWHM is in the range $0.65 - 1.19\arcsec$. We generated a deep stacked image for each area as the superposition of all the epochs with seeing FWHM $< 0.8\arcsec$. Table \ref{tab::epochs} reports the essential information about each epoch in the CDF-S3 and 4.

\begin{table}[ht]
        \small
  \centering
        \renewcommand{\arraystretch}{0.87}
   \begin{tabular}{cccc}
    \toprule
     Epoch ID & Date & Field & FWHM \\
       & (y-m-d) &  & (arcsec) \\
    \toprule
                        779883 & 2013-01-02 & CDF-S3 & 0.84 \\
                        779888 & 2013-01-05 & CDF-S3 & 0.81 \\
                        915014 & 2013-08-01 & CDF-S3 & 0.73 \\
                        915017 & 2013-08-04 & CDF-S3 & 0.83 \\
                        915026 & 2013-08-14 & CDF-S3 & 0.68 \\
                        915029 & 2013-08-16 & CDF-S3 & 0.69 \\
                        915050 & 2013-08-26 & CDF-S3 & 0.87 \\
                        915053 & 2013-08-29 & CDF-S3 & 0.79 \\
                        915059 & 2013-09-04 & CDF-S3 & 0.87 \\
                        915062 & 2013-09-08 & CDF-S3 & 1.02 \\
                        915065 & 2013-09-12 & CDF-S3 & 0.79 \\
                        915074 & 2013-09-24 & CDF-S3 & 0.82 \\
                        915077 & 2013-09-27 & CDF-S3 & 1.01 \\
                        994410 & 2013-10-02 & CDF-S3 & 0.94 \\
                        994424 & 2013-10-04 & CDF-S3 & 1.16 \\
                        994432 & 2013-10-06 & CDF-S3 & 0.87 \\
                        \textbf{994458} & \textbf{2013-10-11} & \textbf{CDF-S3} & \textbf{0.70}  \\
                        994098 & 2013-10-20 & CDF-S3 & 0.72 \\
                        994469 & 2013-10-24 & CDF-S3 & 1.01 \\
                        994111 & 2013-10-25 & CDF-S3 & 0.62 \\
                        994482 & 2013-10-26 & CDF-S3 & 0.56 \\
                        994125 & 2013-10-27 & CDF-S3 & 0.52 \\
                        994490 & 2013-10-29 & CDF-S3 & 0.87 \\
                        994527 & 2013-11-06 & CDF-S3 & 0.88 \\ 
                        994535 & 2013-11-09 & CDF-S3 & 1.01 \\
                        994176 & 2013-12-21 & CDF-S3 & 0.83 \\
                        994188 & 2013-12-23 & CDF-S3 & 0.64 \\
                        994206 & 2013-12-29 & CDF-S3 & 1.20  \\
                        994210 & 2014-01-01 & CDF-S3 & 0.72 \\
                        994225 & 2014-01-05 & CDF-S3 & 0.87 \\
                        994234 & 2014-01-07 & CDF-S3 & 1.01 \\
                        1139063 & 2014-12-14 & CDF-S3 & 0.68 \\
     stacked & & CDF-S3 & 0.76 \\
    \midrule
        994264 & 2013-10-09 & CDF-S4 & 0.73 \\
      1049199 & 2014-07-30 & CDF-S4 & 0.67 \\
      1049203 & 2014-08-02 & CDF-S4 & 0.71 \\
      1049206 & 2014-08-06 & CDF-S4 & 1.19 \\
      1049210 & 2014-08-16 & CDF-S4 & 0.96 \\
      1049225 & 2014-09-01 & CDF-S4 & 0.96 \\
      1049254 & 2014-09-24 & CDF-S4 & 0.82 \\
      1049258 & 2014-09-26 & CDF-S4 & 0.82 \\
      1049262 & 2014-09-29 & CDF-S4 & 1.02 \\
      1049266 & 2014-10-01 & CDF-S4 & 0.65 \\
      1116989 & 2014-10-12 & CDF-S4 & 0.87 \\
      1116998 & 2014-10-19 & CDF-S4 & 0.80 \\
      1117002 & 2014-10-22 & CDF-S4 & 0.90 \\
      1117006 & 2014-10-25 & CDF-S4 & 1.08 \\
      \textbf{1117010} & \textbf{2014-10-28} & \textbf{CDF-S4} & \textbf{0.73} \\
      1144898 & 2014-11-09 & CDF-S4 & 0.78 \\
      1144906 & 2014-12-01 & CDF-S4 & 0.84 \\
      1144911 & 2014-12-04 & CDF-S4 & 0.74 \\
      1117043 & 2014-12-10 & CDF-S4 & 0.99 \\
      1117048 & 2014-12-12 & CDF-S4 & 0.87 \\
      1117052 & 2014-12-15 & CDF-S4 & 1.07 \\
      1117056 & 2014-12-17 & CDF-S4 & 0.77 \\
      1117060 & 2014-12-21 & CDF-S4 & 0.91 \\
      1117064 & 2014-12-23 & CDF-S4 & 0.65 \\
      1117069 & 2015-01-08 & CDF-S4 & 0.85 \\
      1117073 & 2015-01-12 & CDF-S4 & 0.67 \\
      1117077 & 2015-01-16 & CDF-S4 & 0.72 \\
      1145335 & 2015-01-20 & CDF-S4 & 1.00 \\
      1174542 & 2015-01-29 & CDF-S4 & 0.85 \\
      1174544 & 2015-01-31 & CDF-S4 & 0.94 \\
      stacked & & CDF-S4 & 0.78 \\
    \bottomrule
                \end{tabular}
   \caption{\label{tab::epochs}Essential information about each epoch in the CDF-S3 and 4. The bold lines refer to the reference epoch in each region.}
\end{table}

\section{Catalog extraction}
\label{Catalog_extraction}
In the process of identifying variable sources, we adopted an approach similar to that described in \citet{DeCicco}, \citet{decicco19}, and \citet{Falocco}.
First, with the help of \emph{SExtractor} \citep{Bertin96}, we extracted sources from all epochs (using a $5.5 \sigma$ detection threshold) and obtained source catalogs with measured aperture magnitudes. We chose a photometric aperture of $2\arcsec$ diameter, which encloses approximately 70\% of the flux of a point-like source: this is therefore a good compromise in order to obtain a good nuclear flux measurement and low background noise. This aperture also makes our measurements robust to small centering problems in the case of faint AGN in extended sources.

We then used masks to exclude all the sources with possible problems in their photometry; for example, we excluded those sources close to saturated stars or located near the edges of the image, which are typically noisy; or those sources close to regions of the image affected by aesthetic defects. We used the same initial mask for all images: this was obtained making use of the script developed by \citet{Huang} that detects and masks all the stars and their corresponding halos. The mask was then improved manually using the software \emph{SAOImage DS9} to mask additional edge noise and image defects. In total we removed nearly 20\% of the sources in each catalog, which is consistent with the previous studies ($\sim$20\% in both \citealt{Falocco} and \citealt{decicco19}). The single-epoch catalogs were then matched within a radius of $1\arcsec$ to obtain a ``master'' catalog with data for all the epochs in which each source was detected.

\begin{figure}[t]
\begin{center}
        \includegraphics[width=\linewidth]{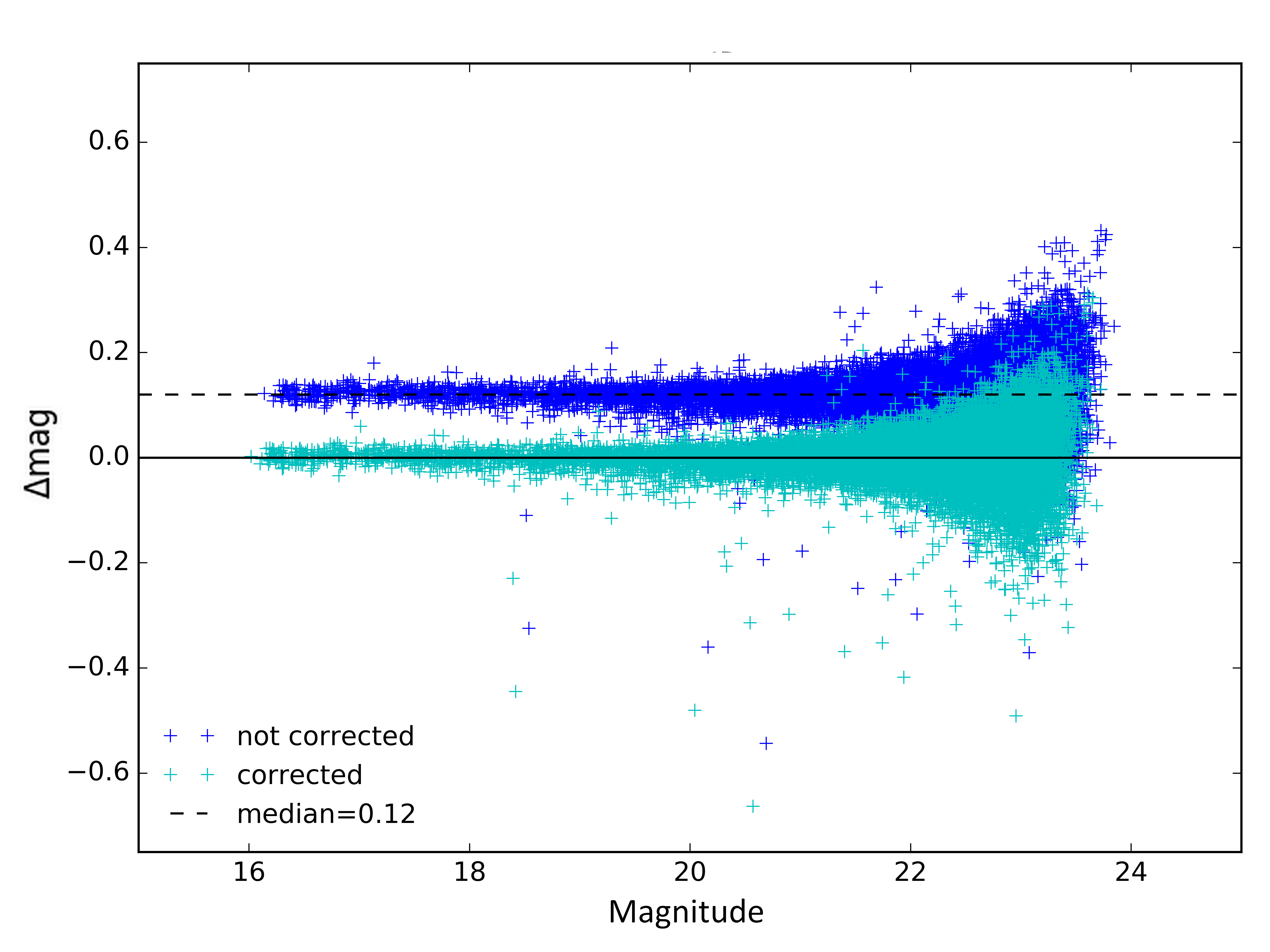}
        \caption{\label{fig::mag_correct}Magnitude difference between epoch 994432 (CDF-S3) and the reference epoch as a function of the magnitude of the sources in the reference epoch. Blue crosses: sources detected in both epochs, before correction. Cyan crosses: sources detected in both epochs, after correction. Dashed line: median of the magnitude difference. Only relatively bright sources, which are detected in all epochs, are shown.}
\end{center}
\end{figure}

\begin{figure*}[ht]
\begin{center}
        \includegraphics[scale=0.55]{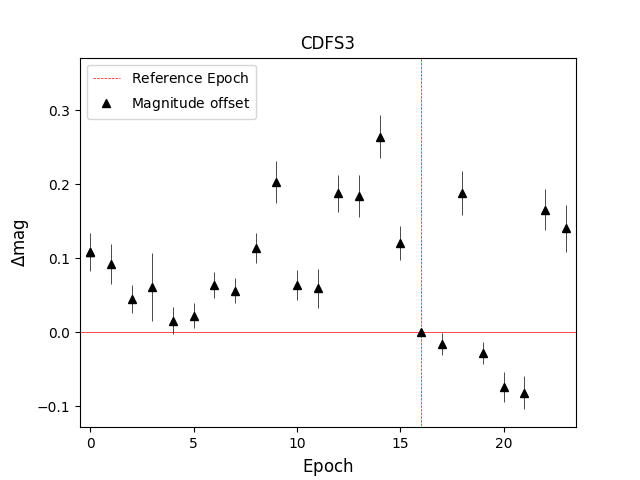}
        \includegraphics[scale=0.55]{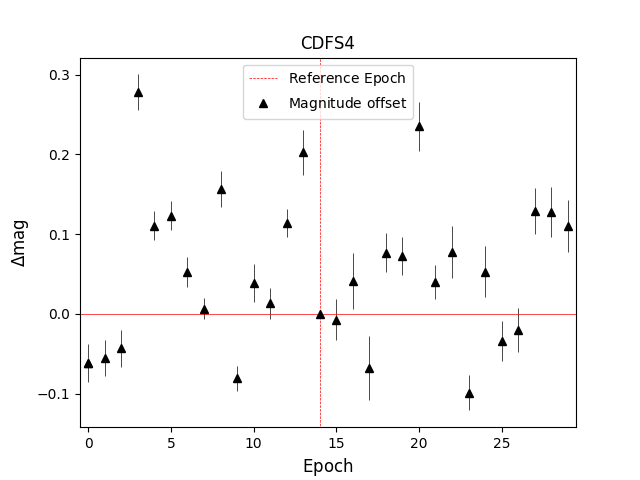}
        \caption{\label{fig::dmag_vs_epochs} Median of the magnitude difference for each epoch in the CDF-S3 (\emph{left panel}) and 4 (\emph{right panel}) taking into account sources < 19 mag. Black triangles: median of the magnitude difference. Red dashed line: reference epoch. Error bars: r.m.s. of the magnitude difference.}
\end{center}
\end{figure*}

Although the VST images are photometrically calibrated by the initial pipeline (VST tube; \citealt{Grado}), residual calibration uncertainties may introduce spurious variability; furthermore our photometry uses 2\arcsec apertures and thus we must correct for point spread function (PSF) variations that affect our flux measurements. We first selected all the sources detected in all epochs (nearly 16000 sources for CDF-S3 and 10000 for CDF-S4)
to estimate the corrections for each epoch. We then chose an epoch to be used as a reference (994458 for CDF-S3 and 1117010 for CDF-S4). We required this reference epoch to have good seeing, i.e., a FWHM of about 0.7\arcsec. Then, for all the sources in each epoch, we plotted the magnitude difference with respect to the reference epoch as a function of the magnitude. The result for one of the epochs is reported in Figure \ref{fig::mag_correct} as an example, while in Figure \ref{fig::dmag_vs_epochs} we show for each epoch the median of the magnitude difference for all the sources with magnitude $\leq$ 19, to avoid taking into account faint sources with low signal-to-noise ratio. It is important to point out that the differences between the epochs is mainly due to PSF variations, while the calibration uncertainties are much smaller. We then used the offset corresponding to each epoch to correct magnitudes. We checked that using a different epoch for CDF-S4 (epoch 1144911) with comparable seeing returns almost identical results, with a difference of $\sim$10-20 sources, most of which have the lowest significance (and thus are more affected by small differences in the adopted variability threshold, which are introduced in next section).
 
\section{Selection of variable sources}
\label{Selection}
Following \citet{DeCicco} and \citet{Falocco} we chose to include in our analysis only sources detected in at least 20\% of the epochs to have robust light curves. This selection criterion returned 29036 sources for the CDF-S3 and 39376 for the CDF-S4\footnote{At this stage the different number of sources between the two fields is due to the different epoch temporal distribution and depth, since no magnitude cut has been applied yet; the magnitude cut is applied later in this section.}. For each source $i$ we derived the average magnitude $\overline{mag_i}$ and the corresponding r.m.s. deviation $\sigma_i$ of the light curve, defined as

\begin{equation}
 \overline{mag_i}=\frac{1}{N_{epoch}}\sum_{j=1}^{N_{epoch}} mag_i^j\mbox{   ,}
\label{eq::mag_mean} 
\end{equation}

\begin{equation}
\sigma_i=\frac{1}{N_{epoch}}\sum_{j=1}^{N_{epoch}} (mag_i^j-\overline{mag_i})\mbox{   ,}
\label{eq::rms}
\end{equation}
where $N_{epoch}$ indicates the number of epochs in which the source was detected.

\begin{figure*}
\begin{center}
        \includegraphics[scale=0.55]{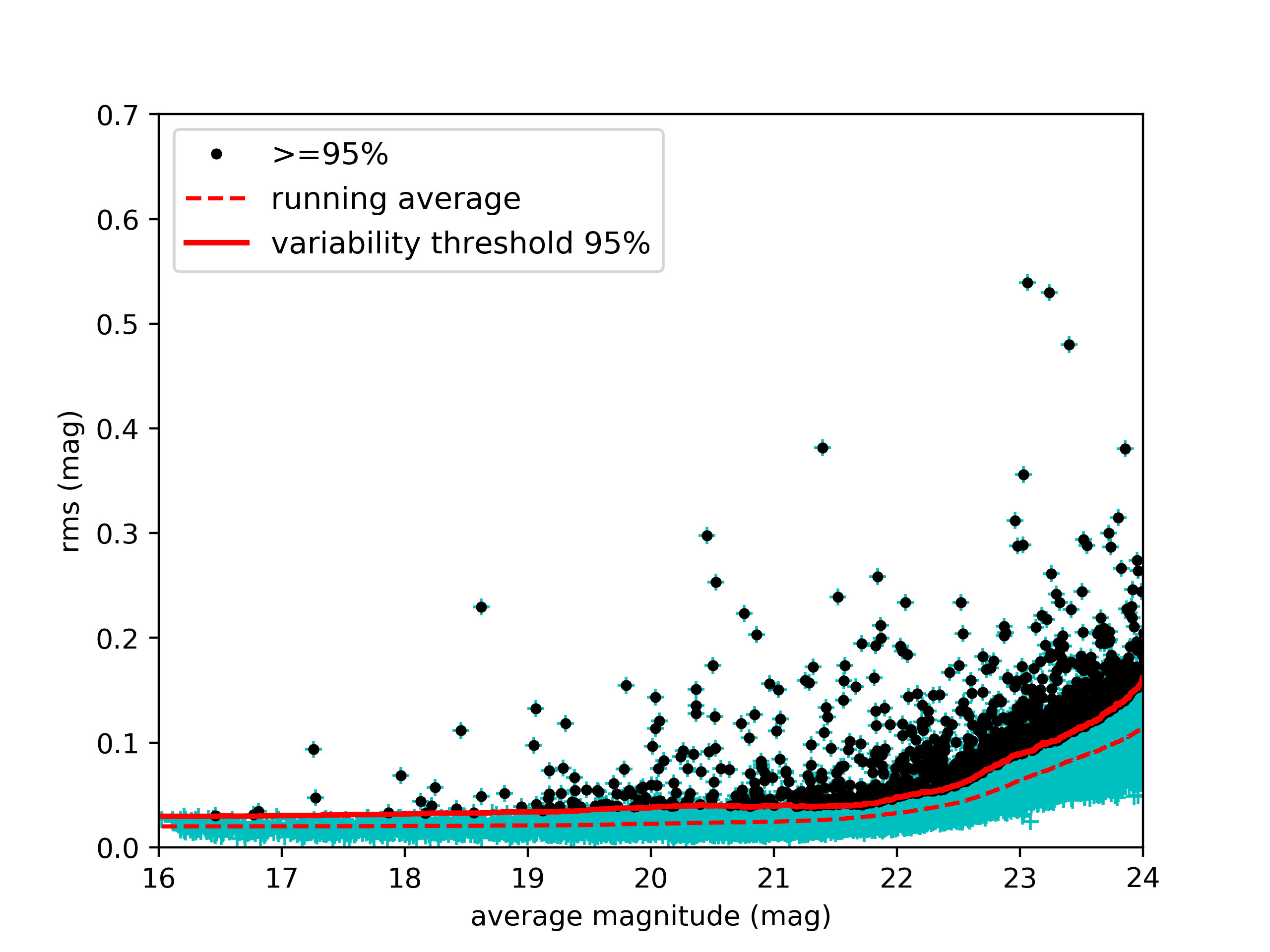}
        \includegraphics[scale=0.55]{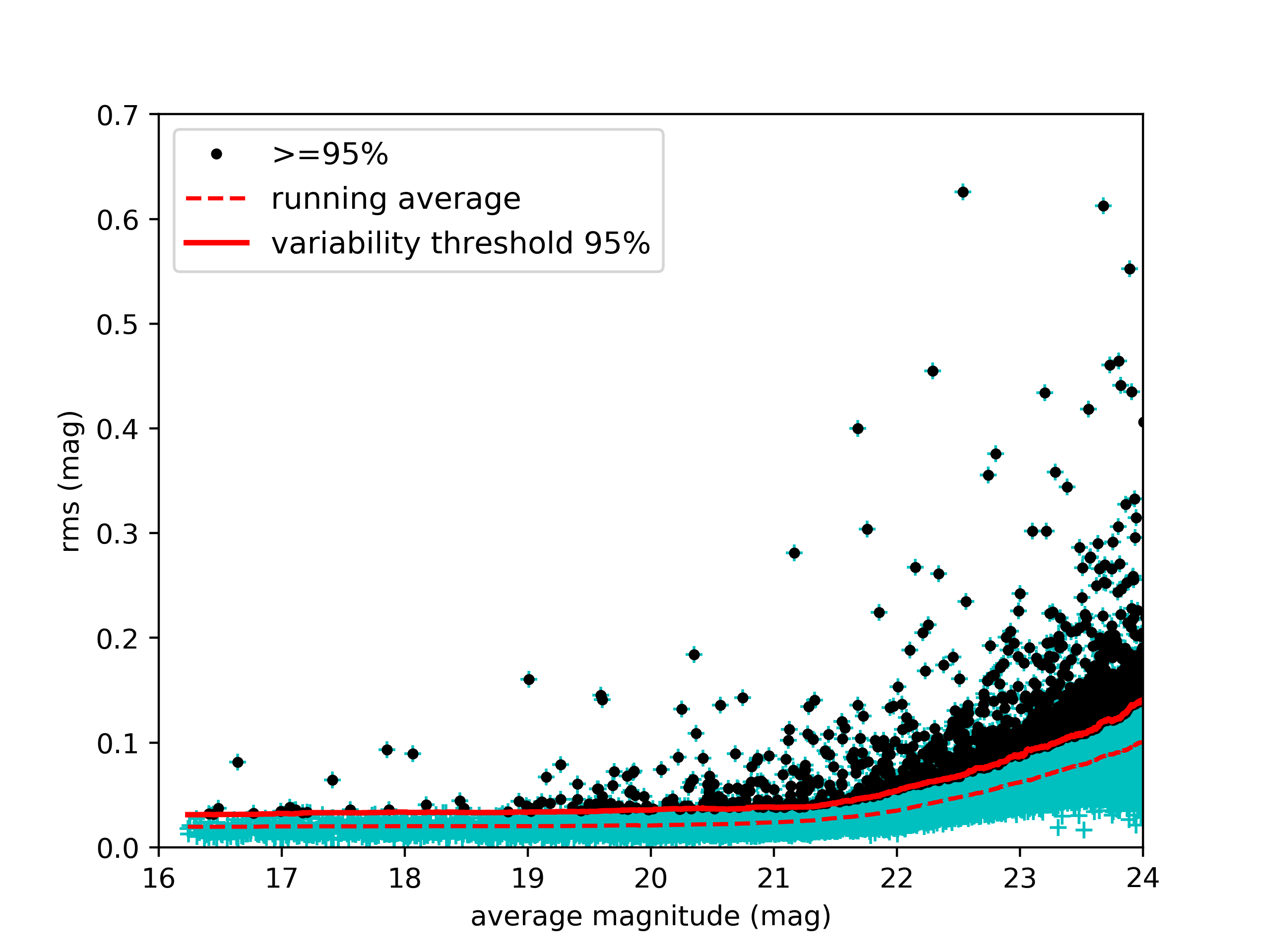}
        \caption{\label{fig::rms_vs_mag}Deviation of r.m.s. as a function of the average magnitude for all the selected sources in the CDF-S3 (\emph{left panel}) and 4 (\emph{right panel}) detected in at least 20\% of epochs. Cyan crosses: selected sources. Solid red line: variability threshold (95th percentile of the r.m.s. distribution). Dashed red line: running average of the r.m.s. deviation. Black points: variables sources (i.e., above the defined variability threshold).}
\end{center}
\end{figure*}

We defined our running variability threshold as in \citet{decicco19}. We considered as variable all the sources having a r.m.s. deviation larger than 95\% with respect to all other sources of similar brightness, i.e., within an interval of $\pm$1000 points centered on the source average magnitudes. The resulting selection is shown in Figure \ref{fig::rms_vs_mag}. 

At faint magnitudes (\emph{r} $>$ 23 mag) we can see that the effect of statistical uncertainties due to low fluxes increases rapidly, so that the variability detection threshold is $>$ 0.05 mag. In agreement with \citet{DeCicco} and \citet{Falocco}, we chose to study only sources with r $\leq$ 23 mag, where the variability threshold is relatively constant. Based on this criterion, our sample is composed of 684 and 675 variable AGN candidates for CDF-S3 and 4, respectively. In any case sources with spurious variability are still be present in our samples of variable sources. In addition to the $<$ 5\% contaminants expected on the basis of our statistical criterion, in extended sources (nearby galaxies) the centroid in which the AGN resides is hard to identify accurately. Thus, we removed such sources by using FLUX\_RADIUS, one of the parameters computed by \emph{SExtractor}. This parameter corresponds to the half-light radius of the source. Knowing that this value is usually lower than 1\arcsec for point-like sources, we decided to exclude from our analysis all the sources with a radius larger than 1.5\arcsec. Finally, we inspected each candidate in every epoch to make sure that they were isolated and not affected by aesthetic problems or close neighbors that could contaminate the photometry. We assigned each source one of the following flags according to their observation quality: 1) a very good candidate with no evidence of problems, 2) a good candidate, but possibly affected by a close companion or minor aesthetic problems, and 3) a bad candidate with likely spurious variability. 

\begin{figure}
\begin{center}
        \includegraphics[width=\linewidth]{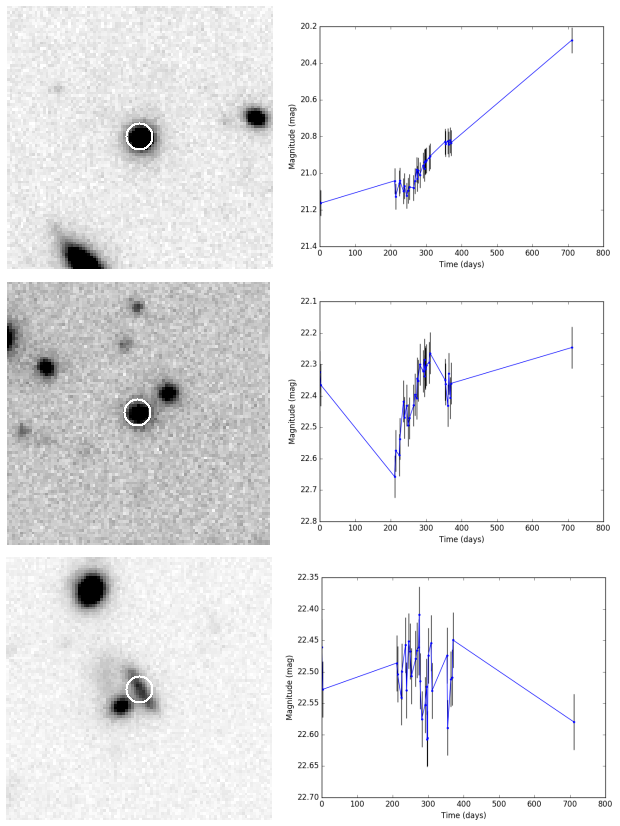}
        \caption{\label{fig::quality_flags}Snapshots of sources taken from the stacked image of CDF-S3 for each quality flags and their corresponding light curve over 23 months (from January 2, 2013 to December 14, 2014). The snapshots have a size of $20\arcsec\times20\arcsec$ and the white circles have a 2\arcsec diameter. Top: source classified as 1 because isolated and without any aesthetic problem. Middle: source classified as 2 because of the close luminous neighbor. Bottom: extended source classified as 3 because of the close and bright companion. Error bars correspond to the r.m.s. of the variability threshold at the average magnitude of the source.}
\end{center}
\end{figure}

An example of a source for each flag can be found in Figure \ref{fig::quality_flags}. We limited our analysis to sources flagged as 1 and 2. We obtained a sample of 350 sources for the CDF-S3 and 432 sources for the CDF-S4. The larger number of variable sources removed from CDF-S3 results from a larger number of extended sources in that region.\\

\section{Nature of variable sources}
\label{Nature}
In this section we investigate the nature of our variable sources using different diagnostics, extending those introduced in \citet{Falocco} and \citet{decicco19}. To produce the various diagnostics, we made use of optical and IR data from the SERVS and SWIRE surveys from the \emph{Spitzer} mission, which provide the optical SDSS \emph{ugriz} magnitudes and the Infrared Array Camera (IRAC) fluxes at 3.6 $\mu$m, 4.5 $\mu$m, 5.8 $\mu$m, and 8.0 $\mu$m. \citet{Vaccari} describes how the two datasets are merged.

        \subsection{Supernova identification}
        \label{SNe_section}
\begin{figure}[htb]
\begin{center}
        \includegraphics[width=\linewidth]{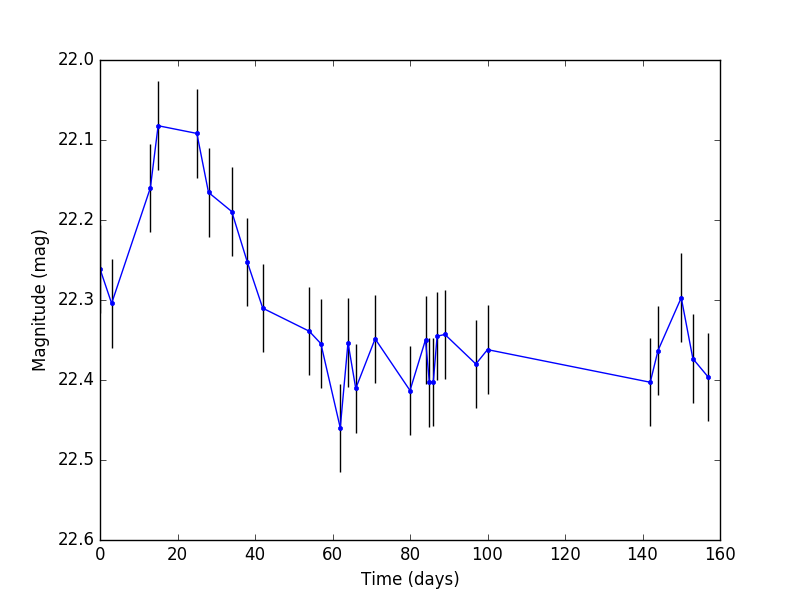}
        \caption{\label{fig::SNe}Light curve over 5 months (from August 1, 2013 to January 7, 2014) of a SN from the CDF-S3 variable sources sample. We cut the light curve to five months instead of 23 for a better visualization of the peak. The SN peaks at day $\sim$20; then it fades, and the only relevant contribution to the luminosity comes from the host galaxy. Error bars refer to the r.m.s. of the variability threshold at the average magnitude of the source.} 
\end{center}
\end{figure}

        Finding SNe is the main goal of the SUDARE survey and therefore we expect to find several SNe among our variable sources. The typical light curve of a SN is characterized by a rapid increase in the luminosity followed by a steep decrease and then a flat part, as shown in Figure \ref{fig::SNe}. As a first step we visually inspected the light curves of our variable sources to identify SNe, yielding 13 and 14 candidates in the CDF-S3 and 4, respectively. We compared the results of our visual inspection to a preliminary list of SNe provided by the SN search team of the SUDARE program (Ragosta et al., in prep.). Their identification procedure is based on spectral template fitting. We could confirm five SNe in the CDF-S3 and seven in the CDF-S4. Similar to \citet{Falocco}, these 12 sources are removed from the sample to compute the final results (see Section \ref{Discussion_conclusion}).
                
We note that the number of SN contaminants is consistent with the findings of \citet{DeCicco} and \citet{Falocco}, since we spanned a baseline of nearly a year. In \citet{decicco19} instead we studied the COSMOS region using a three year baseline, which resulted in a greatly reduced SN contamination rate since such objects vary on shorter timescales and are more
 easily identified with improved sensitivity to long-term variability.
        
        \subsection{Near-infrared/optical diagnostic}
        
\begin{figure}[htb]
\begin{center}
        \includegraphics[width=\linewidth]{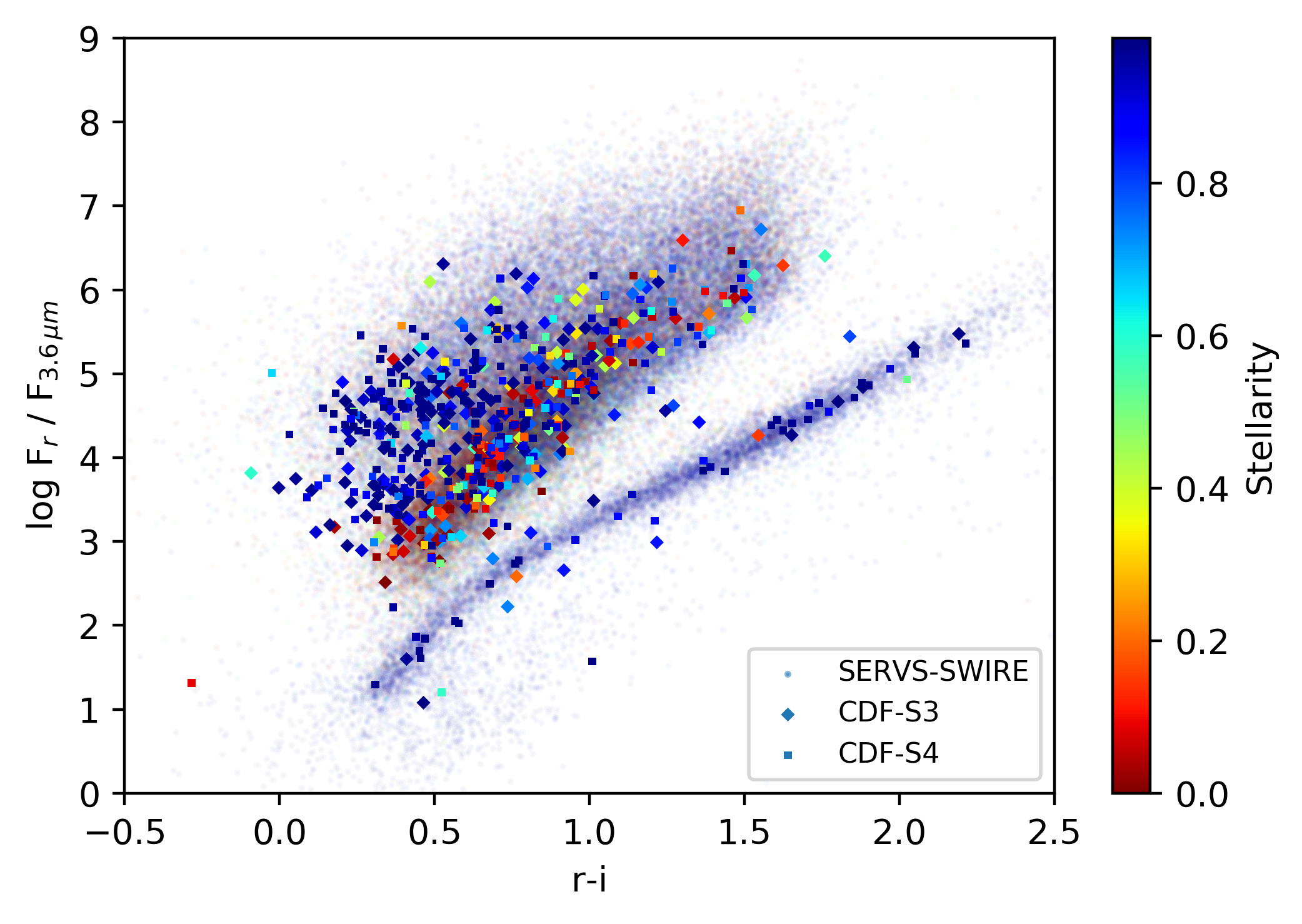}
        \caption{\label{fig::rowan}Near-infrared/optical diagnostic from \citet{Rowan}, representing \emph{r-i} bands vs. log(F$_{r}$/F$_{3.6 \mu m}$). Dots: SERVS-SWIRE reference population with available r, i, 3.6 $\mu$m photometry. Diamonds: CDF-S3 variable sources. Squares: CDF-S4 variable sources. Color bar: stellarity index from 0 (red) for galaxies to 1 (blue) for star-like sources.}
\end{center}
\end{figure}

        To identify stars in our samples of variable sources, we made use of the diagnostic developed by \citet{Rowan}. This allowed us to separate stars from galaxies by comparing the \emph{r-i} color and the ratio of the fluxes measured in the \emph{r} and 3.6 $\mu$m bands. The use of this diagnostic is limited to the variable sources in CDF-S3 and CDF-S4 fields that were detected in these bands (246 and 338 sources, respectively). We show the result for both regions in Figure \ref{fig::rowan}, where we can clearly distinguish two populations: the stars on the bottom line, and the extragalactic sources in the group at the top of the diagram. We find that 17 and 37 sources in the CDF-S3 and 4, respectively, are on the stellar sequence; such numbers correspond to $\sim$ 7\% and $\sim$ 11\% of the variable sources (with NIR/optical data) in each field, in agreement with the findings of \citet{Falocco}.
        
        \subsection{Mid-infrared diagnostic}

        Once we identified SNe and stars within our samples of optically variable sources, we made use of a diagnostic based on IRAC mid-infrared (MIR) colors to find AGN. \citet{Lacy2004} identified a region where AGN are generally located on such diagram, and later  \citet{Donley} proposed a more restrictive selection criterion. An additional AGN region was defined by \citet{Stern} using IRAC bands as well. There are 134 and 155 variable sources with an IR counterpart in all four IRAC bands in the CDF-S3 and CDF-S4 samples, respectively, for which we can cross-check for the presence of a MIR-selected AGN.
                
                \subsubsection{Lacy color diagram}
                
\begin{figure}[t]
\begin{center}
        \includegraphics[width=\linewidth]{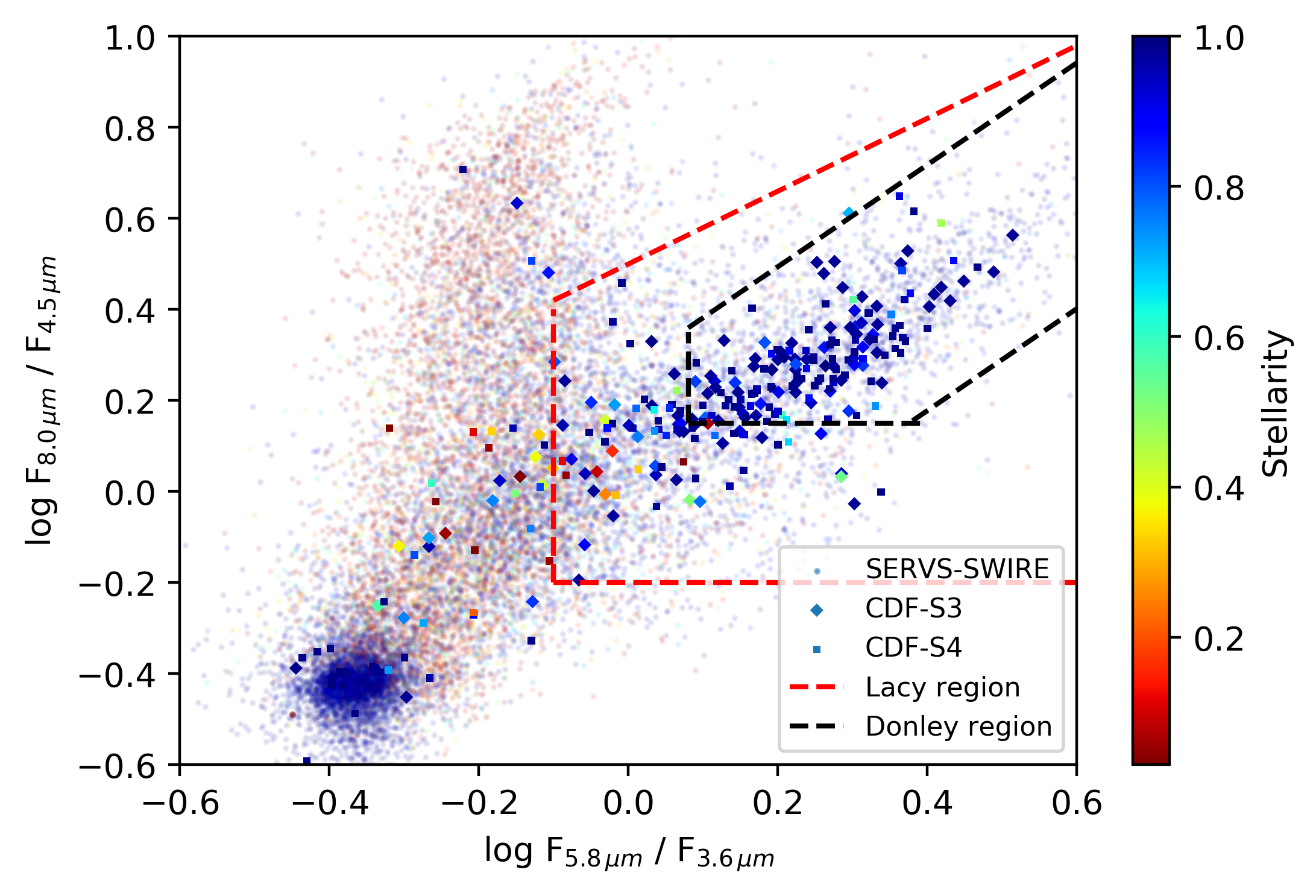}
        \caption{\label{fig::lacy_donley}Mid-infrared diagnostic by \citet{Lacy2004, Lacy2007}. Dots: SERVS-SWIRE reference population. Diamonds: CDF-S3 variable sources. Squares: CDF-S4 variable sources. Color bar: stellarity index from 0 (red) for galaxies to 1 (blue) for star-like sources. Red dashed lines: \citet{Lacy2007} AGN selection criteria. Black dashed lines: \citet{Donley} AGN selection criteria.} 
\end{center}
\end{figure}

                This diagram allows use to distinguish the nature of the sources based on their IR colors by plotting \(y=log(\frac{F_{8.0\mu m}}{F_{4.5\mu m}})\) as a function of \(x=log(\frac{F_{5.8\mu m}}{F_{3.6\mu m}})\). In Figure \ref{fig::lacy_donley}, sources (\(\approx-0.4\)) with bluer colors on both axes are mostly stars or low-redshift galaxies. The rest of the sources occupy two different regions: sources characterized by blue colors on the \emph{x}-axis and  red colors on the \emph{y}-axis mainly correspond to star-forming galaxies, while sources along the diagonal with red colors on both axes are very likely AGN, whose IR emission is due to the presence of nuclear dust.
                To identify the region preferentially occupied by AGN, we adopted the revised criteria from \citet{Lacy2007}, i.e.,

\begin{equation}
\label{eq::lacy_criteria}
\begin{gathered}
 x>-0.1;~ y>-0.2\mbox{;}\\
 y\leq0.8x+0.5\mbox{.}
\end{gathered}
\end{equation}

                 In Figure \ref{fig::lacy_donley} we also show the region identified by the more restrictive criteria of \citet{Donley} as follows:

\begin{equation}
\label{eq::donley_criteria}
\begin{gathered}
 x\geq0.08;~ y\geq0.15\mbox{;}\\
 y\leq1.12x+0.27\mbox{;}\\
 y\geq1.12x-0.27\mbox{.}
\end{gathered}
\end{equation}
                 
                 The Lacy and Donley criteria mainly allow the separation of AGN from star-forming galaxies, which also emit in the IR. The Donley selection criterion relies on the fact that the AGN SED can be approximated by a power-law function \(F_{\nu}=C\nu^{-\alpha}\), \(F_{\nu}\) ; that is, the flux per unit frequency interval expressed in ergs s\(^{-1}\) cm\(^{-2}\) Hz\(^{-1}\), while C is a constant and \(\alpha\geq 0.5\) represents the power-law index. The Donley criteria therefore identify a region occupied by sources with an AGN-dominated SED, and therefore identify highly reliable, although less complete, samples.

It is evident that the majority of our variable sources tend to occupy the Lacy region and tend to align along the power-law strip. In fact we found that 111 out of the 134 sources in the CDF-S3 and 121 out of the 155 sources in the CDF-S4 lie in the Lacy region, while we find 70 CDF-S3 sources and 83 CDF-S4 sources in the Donley region. We notice that one of the CDF-S3 confirmed SNe falls in both regions and, according to Section \ref{SNe_section}, are removed from the final sample.

\citet{Donley} point out that some of the high redshift sources with $z>2.7$ may be star-forming contaminants. These authors define an additional criterion to identify and remove this type of sources. Within our sample of sources falling in Donley wedge and having an available redshift, one source in CDF-S3 and one in CDF-S4 have a redshift > 2.7. These two sources do not satisfy the criterion and thus are unlikely star-forming contaminants. Of the other sources satisfying this additional flux criterium, none of these have a redshift > 2.7, and thus cannot be excluded from our final sample.

                \subsubsection{Stern color diagram}
                
                An alternative diagnostic plot was proposed by \citet{Stern}. This diagnostic plots magnitude differences \(y=[3.6\mu]-[4.5\mu]\) as a function of \(x=[5.8\mu]-[8.0\mu]\) to separate stars, galaxies, and AGN. Star colors on both axes are typically around 0 due to the Rayleigh-Jeans tail of their blackbody spectrum in the MIR. Galaxies occupy different positions depending on their redshift: those with z $\leq$ 1 are mostly situated in the region -0.2 $\lesssim$ y $\lesssim$ 0.5, while galaxies with z $\geq$ 1 shift to redder colors on the same axis, typically with y $\gtrsim$ 0.5. Finally, we have AGN that form a rather vertical branch owing to a lack of strong emission of polycyclic aromatic hydrocarbons\footnote{Organic compounds consisting of carbon and hydrogen that can be found in IR luminous galaxies.} (PAHs). The poor PAH emission, on one hand, restricts AGN observations on the \emph{x}-axis and, on the other hand, makes AGN redder on the \emph{y}-axis because of their blackbody spectrum. \citet{Stern} proposed the following criteria to isolate AGN from galaxies and stars:
\begin{equation}
\label{eq::stern_criteria}
\begin{gathered}
 x>0.6\mbox{;}\\
 y>0.2x+0.18;~ y<2.5x-3.5\mbox{.}
\end{gathered}
\end{equation}

\begin{figure}[t]
\begin{center}
        \includegraphics[width=\linewidth]{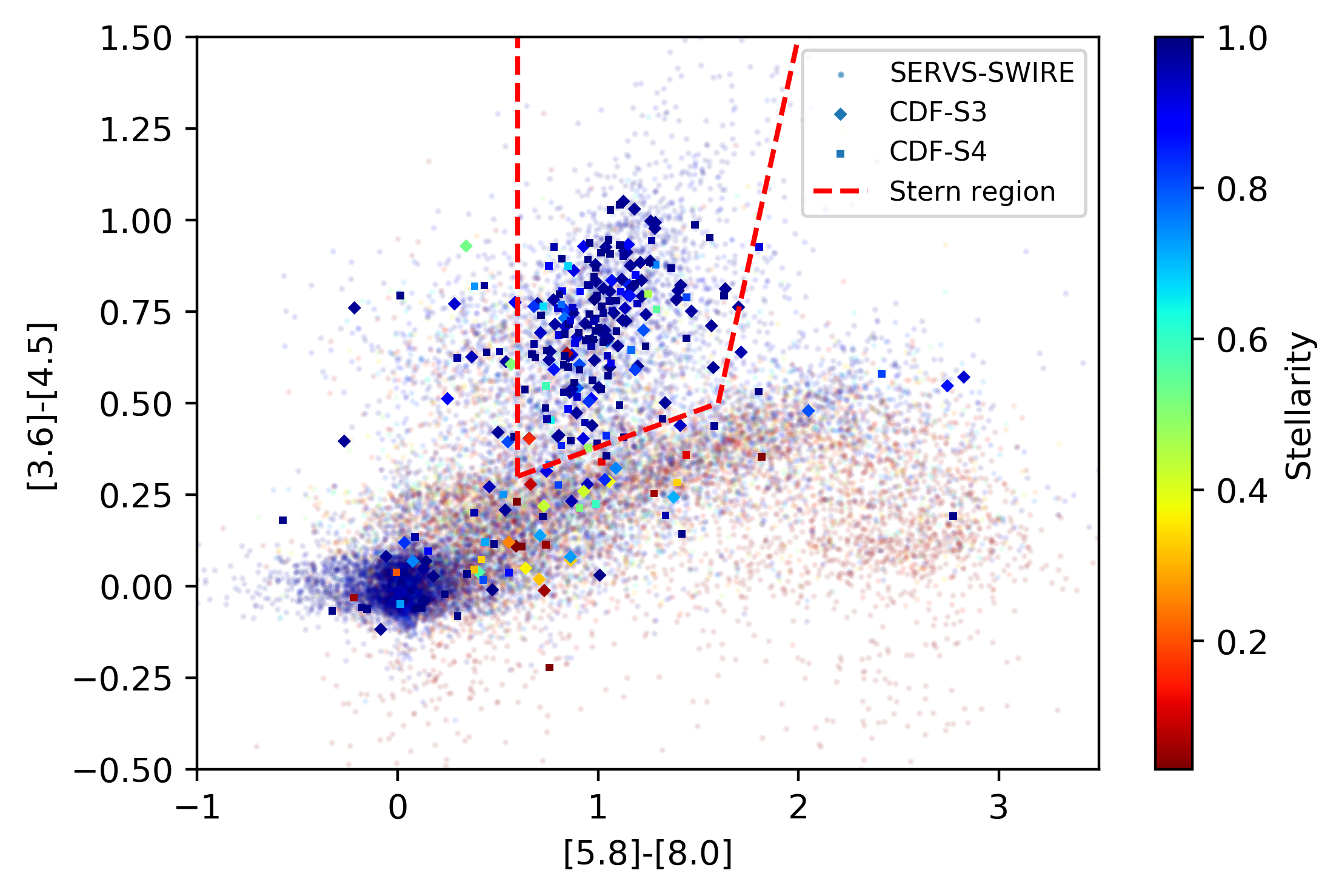}
        \caption{\label{fig::stern_diag}Mid-infrared diagnostic proposed by \citet{Stern}. Dots: SERVS-SWIRE reference population. Diamonds: CDF-S3 variable sources. Squares: CDF-S4 variable sources. Color bar: stellarity index from 0 (red) for galaxies to 1 (blue) for star-like sources. Red dashed lines: \citet{Stern} AGN selection criteria.}
\end{center}
\end{figure}

The Stern diagram for our samples of sources is shown in Figure \ref{fig::stern_diag}. Of the 134 and 155 sources present on the diagram, 87 and 101 are situated in the Stern region and hence are likely AGN. The same confirmed SN lying in the Lacy and Donley regions is also falling in this region and is be taken into account for the final results.

\section{Discussion and conclusions}
\label{Discussion_conclusion}
The effectiveness of our method for selecting AGN based on their variability has been already discussed at length in our previous papers in this series \citep{decicco19,Falocco,DeCicco}, using samples with similar or better multiwavelength coverage (COSMOS, CDF-S1 and 2). Those works presented the completeness and contamination of variability-selected samples with respect to X-ray and IR selected sources, and their dependence on the observing baseline, cadence, magnitude limit, and AGN type. These results are expected to apply to the sample presented in this work as well, given the similarity of the detection method and dataset. For instance our optical variable sample is biased toward unobscured AGN, since obscured AGN are fainter on average and thus their variability is harder to detect owing to the higher (relative) noise; this is discussed in \citet{decicco19}, who find the completeness ratio of unobscured to obscured AGN to be $\sim$5. Thus we only summarize our findings to show the consistency with our previous work in light of some differences in the methodology with respect to,  for example, \citet{Falocco} for the CDF-S1 and 2.

Table \ref{tab::results} reports detailed results for both the MIR diagnostics we used in our analysis. As SNe can be easily distinguished, we excluded those sources we identified from our samples of optically variable AGN candidates. For the Lacy diagnostic, 83\% (110/133, CDF-S3) and 78\% (121/155, CDF-S4) of our variable sources with IRAC counterparts are MIR-selected AGN candidates. These numbers are useful for comparison among different selection methods, although they do not represent a proper purity (confirmed AGN divided by AGN candidates) estimate of the sample considering that the Lacy wedge is contaminated by star-forming galaxies. However if we conservatively assume that all the sources outside the \citet{Lacy2007} region are non-AGN, we obtain an upper limit to the contamination of our sample, since some galaxies outside the Lacy wedge would host low-luminosity AGN as well; with this hypothesis we obtain a contamination of 17\% (23/133) and 22\% (34/155) for the CDF-S3 and CDF-S4, respectively. In any case the combination of variability and IR colors ensures a high confidence in the AGN nature of the selected sources, since the number of contaminants is expected to be $<10\%$ (see the extensive discussion in \citealt{Falocco}).

On the other hand we found that 52\% (69/133) of the CDF-S3 AGN candidates and 54\% (83/155) of the CDF-S4 AGN candidates fall within the Donley wedge. Still taking into account only sources with detection in all IRAC bands and considering all the sources in the Donley region as confirmed AGN (given the very restrictive criterion, see \citealt{Donley}), the ratio of the number of variability-selected AGN candidates to all the sources (variable and nonvariable) that fall in this region yields an average completeness of $\sim$38\% (152/404) over the two fields.

While the performance of variability selection was already compared to MIR selection based on the Lacy and Donley criteria in \citet{Falocco} and \citet{decicco19}, the performance with respect to the Stern diagnostic is discussed in this work for the first time. We find that 65\% (86/133, CDF-S3; 101/155, CDF-S4) of our variable IR-detected sources lie in the Stern region.
Thus the results obtained for the Stern diagnostic have values between those obtained for the Lacy criteria and those for the Donley criteria. This is consistent with what was expected considering that the Lacy diagnostic is the least restrictive and therefore likely the most contaminated by spurious variable sources, while the Donley diagnostic is the most restrictive. 
To further test the reliability of the Stern diagnostic, we compute the ratio of the number of optically variable AGN candidates and the number of all sources that fall outside the Stern region. Since we set a 95\% threshold, if we assume that all our best candidates are variable as a consequence of stochastic photometric fluctuations, we expect $\sim$ 5\% to be spurious AGN. About 2\% of our variable sources, in fact, fall outside the Stern region in both the CDF-S3 and CDF-S4 regions, which is consistent with the sources being mainly spurious detections.

We note that our catalog (available in Table \ref{tab::catalog}), similar to the COSMOS catalog in \citet{decicco19}, provides the detection significance of each source. This thus allows the user to choose the optimal sample, balancing completeness, and contamination based on the scientific application; highly reliable samples can be obtained using higher significance thresholds (as those used in, e.g., \citealt{DeCicco} and \citealt{Falocco}), while more inclusive samples are obtained with our minimal 95\% threshold. For instance to obtain a homogeneous CDFS dataset that is consistent with the \citealt{Falocco} conservative approach used for CDF-S1 and CDF-S2, it would be necessary to select CDF-S3 and CDF-S4 sources above a variability significance threshold of 99\%.

In \citet{Falocco} a comparison with X-ray detected samples was possible only for the part of the CDF-S1 field overlapping with the ECDFS, since X-ray data were missing for the rest of the region. However in an effort to acquire ancillary data before the beginning of LSST operations, we started an XMM campaign to cover the full W-CDF-S area (Ni Q. et al, in preparation). The data collection is currently still underway; once complete, the new X-ray catalog will allow a more extensive comparison with X-ray selected samples as done for the COSMOS field in \citet{decicco19}. In this context new improved SED and redshift estimates for all sources will be made available and will allow a better understanding of the nature of AGN candidates. A full characterization of the population of variability-selected AGN is of paramount importance for the exploitation of the next generation synoptic surveys, such as those that will be provided by LSST, where millions of AGN will be selected based on both colors and variability; but multiwavelength and spectroscopic data will be available only for a small minority of the candidates.

\begin{acknowledgements}
We acknowledge the financial contribution from the agreement ASI-INAF n.2017-14-H.O (M.P. and F.V.), the CONICYT grants Basal-CATA Basal AFB-170002 (D.D., F.E.B.), the Ministry of Economy, Development, and Tourism's Millennium Science Initiative through grant IC120009, awarded to The Millennium Institute of Astrophysics, MAS (D.D., F.E.B., G.P.). M.V. acknowledges support from the Italian Ministry of Foreign Affairs and International Cooperation (MAECI Grant Number ZA18GR02) and the South African Department of Science and Technology's National Research Foundation (DST-NRF Grant Number 113121) as part of the ISARP RADIOSKY2020 Joint Research Scheme.
\end{acknowledgements}

\renewcommand{\arraystretch}{1.3}

\begin{table*}[htb]
\begin{center}
\begin{tabular}{|c| c| c| c| c| c| c|}
   \hline
    & \multicolumn{2}{c|}{Lacy region} & \multicolumn{2}{c|}{Donley region} & \multicolumn{2}{c|}{Stern region}\\
   \hline\hline
   Region & CDF-S3 & CDF-S4 & CDF-S3 & CDF-S4 & CDF-S3 & CDF-S4 \\
   \hline\hline
   Variable AGN candidates & 83\% & 78\% & 52\% & 54\% & 65\% & 65\% \\
         with MIR confirmation & (110/133) & (121/155) & (69/133) & (83/155) & (86/133) & (101/155) \\
         \hline
         Variable MIR-selected & 19\% & 20\% & 41\% & 35\% & 33\% & 30\% \\
          AGN & (110/582) & (121/608) & (69/167) & (83/237) & (86/257) & (101/342)\\
   \hline
\end{tabular}
\end{center}
\caption{\label{tab::results}Mid-infrared diagnostics results for non-SNe sources.}
\end{table*}

\bibliographystyle{aa}
\bibliography{biblio}

\clearpage
\onecolumn
\renewcommand{\arraystretch}{1.1}
\begin{landscape}
\begin{threeparttable}
\begin{longtable}{c c c c c c c c c c}
\caption{List of the optically variable sources in the sample. Column meanings: (1): field; (2): identification number; (3) and (4): right ascension and declination (J2000); (5): average VST \emph{r}(AB) magnitude; (6): r.m.s. of the light curve; (7): percentile: the number indicates that the $\sigma^{lc}$ of the source is above the threshold of the corresponding percentile; (8): spectroscopic redshift: when available, from SERVS-SWIRE (see \emph{http://www.mattiavaccari.net/df/specz/} for more information, \citealt{Vaccari}); (9): \emph{SExtractor} stellarity index; (10) Classification of the source, corresponding to the sum of the following indices: -2 = confirmed supernova; -1 = star from NIR/optical diagnostic; 0 = no available classification; 1 = AGN from MIR diagnostic, Lacy region; 2 = AGN from MIR diagnostic, Stern region; 3 = AGN from MIR diagnostic, Donley region.}\label{tab::catalog}\\
\toprule field & source ID & RA J2000 (deg) & Dec J2000 (deg) & avg $r(AB)$ mag (mag) & ltc r.m.s. (mag) & percentile & redshift & stellarity & classification\\
(1) & (2) & (3) & (4) & (5) & (6) & (7) & (8) & (9) & (10)\\
\endfirsthead
\caption{Continued.} \\
\toprule field & source ID & RA J2000 (deg) & Dec J2000 (deg) & avg $r(AB)$ mag (mag) & ltc r.m.s. (mag) & percentile & redshift & stellarity & classification\\
(1) & (2) & (3) & (4) & (5) & (6) & (7) & (8) & (9) & (10)\\
\toprule
\endhead
\toprule
\endfoot
\toprule
\ CDF-S3 & 1 & 52.8213 & -29.0572 & 20.81 & 0.04 & 95 & 0.18 & 0.02 & 0\\
\ CDF-S3 & 2 & 51.7462 & -29.0533 & 20.06 & 0.08 & 98 & - & 0.67 & 3\\
\ CDF-S3 & 3 & 51.8300 & -29.0521 & 21.30 & 0.06 & 98 & 0.28 & 0.02 & 0\\
\ CDF-S3 & 4 & 52.3881 & -29.0526 & 22.90 & 0.09 & 96 & 0.88 & 0.14 & 0\\
\ CDF-S3 & 5 & 52.3712 & -29.0525 & 21.94 & 0.12 & 99 & 1.77 & 0.77 & 6\\
\ CDF-S3 & 6 & 52.4138 & -29.0521 & 22.53 & 0.06 & 95 & - & 0.30 & 0\\
\ CDF-S3 & 7 & 52.6589 & -29.0492 & 17.26 & 0.09 & 99 & - & 0.93 & -1\\
\ CDF-S3 & 8 & 51.7250 & -29.0450 & 22.66 & 0.08 & 96 & - & 0.56 & 6\\
\ CDF-S3 & 9 & 51.9282 & -29.0447 & 21.69 & 0.08 & 98 & 1.90 & 0.72 & 6\\
\ CDF-S3 & 10 & 52.6606 & -29.0445 & 22.05 & 0.08 & 98 & 2.04 & 0.75 & 1\\
\ CDF-S3 & 11 & 52.3498 & -29.0438 & 22.57 & 0.07 & 96 & 0.55 & 0.11 & 0\\
\ CDF-S3 & 12 & 52.3691 & -29.0427 & 20.52 & 0.09 & 99 & - & 0.91 & -1\\
\ CDF-S3 & 13 & 51.8301 & -29.0394 & 22.45 & 0.06 & 96 & 0.48 & 0.04 & 0\\
\ CDF-S3 & 14 & 51.8361 & -29.0337 & 21.69 & 0.04 & 96 & 0.43 & 0.02 & 0\\
\ CDF-S3 & 15 & 51.8390 & -29.0249 & 22.45 & 0.07 & 97 & 0.45 & 0.03 & 0\\
\ ... & ... & ... & ... & ... & ... & ... & ... & ... & ...\\
\ CDF-S4 & 1 & 52.8496 & -29.0539 & 22.61 & 0.07 & 95 & 0.54 & 0.14 & 0\\
\ CDF-S4 & 2 & 52.9218 & -29.0454 & 18.07 & 0.09 & 99 & - & 0.89 & -1\\
\ CDF-S4 & 3 & 53.9131 & -29.0422 & 17.17 & 0.03 & 95 & - & 0.92 & -1\\
\ CDF-S4 & 4 & 52.9117 & -29.0423 & 22.15 & 0.07 & 97 & 0.47 & 0.05 & 0\\
\ CDF-S4 & 5 & 53.4666 & -29.0399 & 21.00 & 0.06 & 97 & 1.68 & 0.95 & 6\\
\ CDF-S4 & 6 & 53.7648 & -29.0295 & 22.63 & 0.07 & 95 & 0.48 & 0.20 & 0\\
\ CDF-S4 & 7 & 53.6739 & -29.0263 & 22.76 & 0.16 & 99 & 1.36 & 0.62 & 6\\
\ CDF-S4 & 8 & 53.7325 & -29.0256 & 21.84 & 0.10 & 99 & - & 0.45 & 1\\
\ CDF-S4 & 9 & 53.8208 & -29.0240 & 22.93 & 0.09 & 96 & 0.81 & 0.07 & 0\\
\ CDF-S4 & 10 & 53.1027 & -29.0238 & 19.87 & 0.05 & 98 & - & 0.03 & 0\\
\ CDF-S4 & 11 & 53.8504 & -29.0109 & 22.58 & 0.12 & 99 & 0.67 & 0.69 & 0\\
\ CDF-S4 & 12 & 53.9605 & -29.0029 & 20.86 & 0.04 & 96 & 0.33 & 0.02 & 0\\
\ CDF-S4 & 13 & 53.7986 & -29.0082 & 22.58 & 0.10 & 98 & - & 0.58 & 6\\
\ CDF-S4 & 14 & 53.9548 & -29.0031 & 21.30 & 0.06 & 98 & - & 0.85 & 4\\
\ CDF-S4 & 15 & 53.4613 & -29.0008 & 22.68 & 0.08 & 95 & - & 0.09 & 0\\
\ ... & ... & ... & ... & ... & ... & ... & ... & ... & ...\\
\bottomrule
\end{longtable}
\begin{tablenotes}
      \small
      \item Table \ref{tab::catalog} is available in its entirety in the electronic version of the \emph{Astronomy \& Astrophysics} journal.
    \end{tablenotes}
  \end{threeparttable}
\end{landscape}

\end{document}